\documentclass{elsart}

\usepackage{graphicx}

\usepackage{epsfig}

\usepackage{amssymb}

\begin{document}

\begin{frontmatter}

\title{Probability and complex quantum trajectories: Finding the missing links}

\author{Moncy V. John}

\address{Department of Physics, St. Thomas College, Kozhencherry, Kerala 689641,\\
India.
}

\begin{abstract}
It is shown that a  normalisable probability density can be defined for the entire complex plane in the modified de Broglie-Bohm quantum mechanics, which gives complex quantum trajectories.  This work is in continuation of a previous one that defined a conserved probability for most of the regions in the complex space in terms of a trajectory integral, indicating a dynamical origin of quantum probability. There it was also shown that the   quantum trajectories obtained  are the same characteristic curves that propagate information about the conserved probability density. Though the probability density we now adopt for those regions left out in the previous work is not conserved locally,  the net source of probability for such regions is seen to be zero in the example considered, allowing to make the total probability conserved. The new combined probability density  agrees with the Born's probability everywhere on the real line, as required. A major fall out of the present scheme is that it explains why   in the classical limit the imaginary parts of  trajectories are not observed even indirectly  and  particles are confined close to the real line.

\end{abstract}


\end{frontmatter}

\section{Introduction}

Complex quantum trajectories were first  obtained  \cite{mvj1} by modifying the de Broglie's guiding wave approach to quantum mechanics. Here,  trajectories were  drawn  for the cases of  harmonic oscillator, potential step, wave packets etc.  For getting this trajectory representation, first we  substitute $\Psi=e^{i\hat{S}/\hbar}$ in the Schrodinger equation, which gives the quantum Hamilton-Jacobi equation (QHJE) \cite{oldpaps,goldstein}

\begin{equation}
\frac {\partial \hat{S}}{\partial t} + \left[ \frac{1}{2m}\left( \frac
{\partial  \hat{S}}{\partial x}\right)^2 +V\right]  =
\frac{i\hbar}{2m} \frac{\partial^2 \hat{S}}{\partial x^2}. \label{eq:qhje}
 \end{equation}
Then we postulate an  equation of motion for the particle, similar to that used by de Broglie:

\begin{equation}
m\dot{x} \equiv \frac {\partial \hat{S}}{\partial x}= \frac {\hbar
}{i} \frac {1}{\Psi}\frac {\partial \Psi}{\partial x}.
\label{eq:xdot}   \end{equation}
  The quantum trajectories $x(t)$   were found by  integrating this equation with respect to time; in general, they  lie in a complex $x$-plane, with $x=x_r+i  x_i$. This results in a modified version of the  de Broglie-Bohm (dBB) quantum mechanics  \cite{dbpap,bohm,dBB,valentini}. Eq. (\ref{eq:xdot}) was used by Leacock and Padgett \cite{leacock} to obtain eigenvalues in many bound state problems, without  having to solve the corresponding Schrodinger equation.

It shall be noted that the canonical momentum is not always the mechanical momentum, even in classical mechanics with Cartesian coordinates. For instance, when there are velocity-dependent potentials, the canonical and mechanical momenta are different. In the case of a charged particle in a magnetic field with the vector potential ${\bf A}$ , the mechanical momentum may be written as

\begin{equation}
m{\bf \dot{x}} = {\bf \nabla}{\hat S}-e{\bf A}/c. 
\end{equation}

This equation in Cartesian coordinates shall thus be the equation of motion we adopt for charged particles in higher dimensions with electromagnetic field.

The Floyd-Faraggi-Matone (FFM) trajectory representation \cite{floyd,faraggi,carroll,floyd2,floyd3} is  another modified dBB version but with real trajectories and is based on a generalised Hamilton-Jacobi equation equivalent to that used in dBB. But this representation differs from dBB quantum mechanics mainly in the use of the equation of motion. Here, for stationarity  the equation of motion for the trajectory time $t$, relative to its constant coordinate $\tau$, is given as a function of $x$ by

\begin{equation}
t-\tau =\partial W/\partial E
\end{equation}
where $W$ is the Hamilton's characteristic function appearing in the generalised Hamilton-Jacobi equation and $E$ is the energy. Carroll \cite{carroll} finds that for stationarity the above Jacobi's theorem is valid, for $W$ is a Legendre transform of Hamilton's principal function. Floyd \cite{floyd2} notes that as Jacobi's theorem also determines the equation of motion in classical mechanics, it is universal transcending across the division between classical and quantum mechanics. In this way, FFM  claims to be a  deterministic theory.

The complex function $\hat{S}$, which may be called the complex Hamilton's principal function  in the   present modified de Broglian mechanics, and the complex QHJE (\ref{eq:qhje}) itself are quite different from the corresponding entities in dBB or FFM representations. 
Moreover,  probability is an integral part of the new formulation and one  can also consider equation (\ref{eq:xdot}) as propagating probability densities. In fact, it was shown in a previous work \cite{mvj2} that the   quantum trajectories obtained in this scheme are the same as the characteristic curves propagating information about a conserved probability density. Hence we continue to use (\ref{eq:xdot}), with appropriate modifications as mentioned above when necessary, as the equation of motion in this modified dBB formulation.

 The complex  trajectory approach gives the  paths in the $n=1$ harmonic oscillator  as shown in Fig. 1. In this case, the paths are given by $(\alpha^2x_r^2-\alpha^2x_i^2-1)^2+4\alpha^4x_r^2x_i^2 = \;  A^2$, a constant for each path. Here $A$ is real and positive, and $\alpha^2$ has the usual definition, equal to $m\omega_0/\hbar$.  It is interesting to note that these figures are the famous Cassinian ovals. The Jacobi lemniscate, which is the special case of these ovals,  corresponds to $A=1$.

\begin{figure}[ht] 
\centering{\resizebox {0.5 \textwidth} {0.3 \textheight }  
{\includegraphics {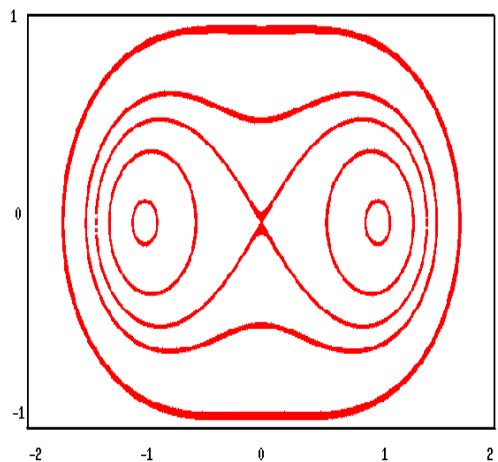}} 
\caption{ The complex trajectories in the $n=1$ harmonic oscillator case, drawn for various values of $A$  and with axes  $X_r\equiv \alpha x_r$ and $X_i\equiv \alpha x_i$. These are the famous Cassinian ovals. The lemniscate corresponds to $A=1$. }  \label{fig:n1shmtraj}} 
   \end{figure}  

  Recently,  the prospects of a dynamical explanation of quantum probability in this scenario was explored  \cite{mvj2}. We have shown that in this modified de Broglian mechanics, the Born's probability along the real axis can be obtained as an exponential function of a line integral along the real axis. Similarly in the extended   $x_rx_i$-plane, a probability density  is obtained as the exponential of a trajectory integral. In both these cases,  integrands involve the particle's complex velocities, indicating a dynamical origin of quantum probability. In the  extended case, we obtain a conserved probability, which agrees with the boundary condition (Born's rule) along the real axis in most regions. However, for other regions (e.g., the subnests inside the lemniscate in the $n=1$ harmonic oscillator case, which do not enclose any poles of $\dot{x}$), such a definition could not be obtained. In fact, a probability density was not proposed for such regions in \cite{mvj2} and it was considered a missing link in the formalism.

Another important question faced by this trajectory representation  is why the complex extended motion is not appreciable in the classical limit. Our everyday experience is that particle trajectories are  in  real space, obeying classical rules. In this letter, we study the probability distribution of particles in the complex plane in detail and adopt for those  regions left out in \cite{mvj2}  an alternative definition, first proposed in \cite{wyatt3,yang3} as the probability density  for the entire plane. This  density satisfies the boundary condition on the real axis, though it is not a conserved one. We  show that the combined probability for the entire extended plane does not diverge and there is no difficulty in normalizing it, contrary to what happens in the above case. Nonconservation of probability points to the prospects of particle creation and destruction, a fundamental feature of quantum phenomena.  Most importantly, the new scheme reveals that large part of the probability for a classical harmonic oscillator lies extremely close to the real line, thereby  explaining why the complex extended motion does not leave any imprint in the classical limit. 

We restrict ourselves to one particle stationary states in 1-dimension for simplicity. The paper  is organized as follows. In the next section, we review the problem with probability density faced in \cite{mvj2}. The third section presents the alternative definition of probability density adopted for the subnests and obtains a form of continuity equation obeyed by it. In section 4, the classical limit of a harmonic oscillator in the light of our combined probability density  and its  implication for complex trajectories are discussed. Section 5 comprises our conclusion.

\section{Probability from velocity field}

As in the case of dBB quantum mechanics, the present modified de Broglie-Bohm mechanics too is constructed in such a way that it agrees with the results of standard quantum mechanics \cite{mvj1}. This is achieved  by accepting the Born's probability axiom along the real line in these representations. However, we note that one of the  challenges before such a quantum trajectory representation  is to explain this probability axiom. In the standard dBB approach, there were several attempts to obtain the $\Psi^{\star}\Psi$ probability distribution from more fundamental assumptions \cite{dBB}. (The FFM trajectory representation does not involve  probability and is considered a  deterministic description.)

 In addition to that defined   along the real line, it is desirable to have a probability density defined everywhere in the extended complex $x$-plane. An earlier attempt made in \cite{poirier} to define such a density  was to write  $\rho (x)= \bar{\Psi (x)}\Psi (x)$, where $\bar{\Psi (x)} \equiv \Psi^{\star}(x^{\star})$, with $x$ complex.  With the help of time-dependent Schrodinger equation, the author shows that, in general, ${\partial \rho}/{\partial t}\neq j^{\prime}(x,t)$.
This arguably leads to nonconservation of probability along trajectories. But it shall be noted that this negative result is based  on the choices made in \cite{poirier} for the probability density and flux. Moreover, this definition leads to complex probability off the real axis, which is  undesirable. 
Another proposal is to define probability as $\Psi^{\star}(x)\Psi(x)$ itself \cite{wyatt3,yang3}. Though this has the advantage of being  real  everywhere, it is not  conserved at any point in the extended plane.  In addition, this  diverges for large values of $x_i$ in many cases and is not generally a normalisable  probability.

 Using a  different approach, it was shown   in \cite{mvj2} that the complex trajectory representation is capable of explaining  the  quantum probability as originating from dynamics itself. Here, the Born's probability density to find the particle on the real axis around some point at $x=x_{r0}$  was  obtained as

\begin{equation}
\Psi^{\star}\Psi (x_{r0},0,t) \equiv P(x_{r0},t) ={\cal N} \exp \left({-\frac{2m}{\hbar}\int^{{x_{r0}}} \dot{x}_i dx_r}\right) , \label{eq:psistarpsi}
\end{equation}
where the integral is taken along the real line.  In addition, an extended probability density $\rho(x_r,x_i)$  in the $x_rx_i$-plane for stationary states was  proposed in \cite{mvj2} and showed, with the aid of complex-extended Schrodinger equation, that the continuity equation follows from it.  The proposal was that if $\rho_0$, the extended probability density at some point $(x_{r0},x_{i0})$ is given, then $\rho(x_r,x_i)$ at another point that lies on the trajectory which passes through $(x_{r0},x_{i0})$ is

\begin{equation}
\rho (x_r(t),x_i(t)) = \rho_0  \exp\left[  \frac{-4}{\hbar}\int_{t_0}^{t} Im\left(\frac{1}{2}m\dot{x}^2+V(x)\right)dt^{\prime}    \right]. \label{eq:rho_def}
\end{equation}
Here, the integral is taken along the trajectory $[x_r(t^{\prime}),x_i(t^{\prime})]$ which passes through $(x_{r0},x_{i0})$. The continuity equation 
 was shown to follow from this axiom by using the extended version of the Schrodinger equation. While evaluating $\rho$ with the help of (\ref{eq:rho_def}) above, one should specify $\rho_0$ at $(x_{r0},x_{i0})$ and if we choose this point as $(x_{r0},0)$, the point of crossing of the trajectory on the real line, then $\rho_0$ can assume the Born's value $P(x_{r0})$.

However, we may note that in some regions of the complex plane, there can be disagreement between the values of $\rho$ and $P(x_r)$ at the points of crossing of the trajectories on the real line. This happens  where the complex trajectories do not enclose any poles of $\dot{x}$, described as `subnests' in \cite{mvj1}. Stated more clearly, the problem here is that as a particle trajectory is traversed in these regions, even while the probability $\rho $ at one point of crossing  $x_{r0}$ on the real axis agrees with $P(x_{r0})$,  at the other point, say the point $x_{r1}$ where the trajectory again crosses the real line, the probability calculated according to (\ref{eq:rho_def}) happens to be different from that of $P(x_{r1})$.

On the other hand, if we directly solve the continuity equation  for  stationary states, it is easy to see that  its solution and the probability density given by the trajectory integral (\ref{eq:rho_def}) give identical results for regions outside the subnests \cite{mvj2}. But   for the subnests, the boundary condition overdetermines the problem and we are unable to find a solution that agrees with the Born's rule everywhere on  the real   line; i.e., there is complete agreement between the extended probability density $\rho$ and the Born's probability density only in the regions outside the subnests.

\section{Probability inside the subnests}
 
Given this situation,  one can ask whether it is possible to find a  probability density for the subnests that can agree with the Born's rule on the real line, even if it is not conserved in the  region. A natural choice for such a definition is the  extended probability $\rho^{\prime} =\Psi^{\star}(x)\Psi (x)$, suggested in \cite{wyatt3,yang3}.  A trajectory integral form for $\rho^{\prime}$,  similar to that in Eq. (\ref{eq:rho_def}) above, was proposed in \cite{wyatt4}. For stationary states, this can be written as

\begin{equation}
\rho^{\prime} (x_r,x_i) \propto \rho_0 \exp\left[  \frac{-4}{\hbar}\int_{t_0}^{t} Im\left(\frac{1}{2}m\dot{x}^2\right)dt^{\prime}    \right]. \label{eq:rho_alt_ls1}
\end{equation}
We propose to adopt this as the probability density for such subnests. Here also, the integral  is to be evaluated  over the trajectory of the particle. The absence of the potential term $V(x)$ in the integrand  marks this definition from that in (\ref{eq:rho_def}). Therefore this distribution will not be conserved. Instead of the continuity equation, we get

\begin{equation}
\frac{\partial \rho^{\prime}}{\partial t} + \frac{\partial (\rho^{\prime} \dot{x}_r)}{\partial x_r}+\frac{\partial (\rho^{\prime} \dot{x}_i)}{\partial x_i}=\frac{4}{\hbar} \rho^{\prime} (x_r,x_i)\; {\hbox {Im}}[V(x)] \label{eq:noncons_gen}
\end{equation}
Thus there are sources and sinks for probability in the subnests. However, as  seen below, the net source of probability in this region for the $n=1$ harmonic oscillator is zero, indicating that the total probability for the  subnested region can remain conserved. 

Thus instead of being `nonviable', the trajectories in this region reveal an important feature of quantum phenomena, namely nonconservation of particles. This indicates  creation and destruction of particles in such regions. Hence we can anticipate that the representation is capable of allowing a smooth transition to quantum field theory. It is interesting to note that the Floyd-Farraggi-Matone trajectory representation too allows creation and destruction of particles \cite{floyd3}.

\section{Probability and classical limit}

 The extended probability density for the $n=1$ harmonic oscillator in the region inside the lemniscate (for $A<1$ and $x_r>0$), computed using the trajectory integral approach in (\ref{eq:rho_alt_ls1}) is shown overlapped with the `leaf-shaped' surface $\Psi^{\star}\Psi$ in this case,  in Fig. 2.

\begin{figure}[ht] 
\centering{\resizebox {0.8 \textwidth} {0.5 \textheight }  
{\includegraphics {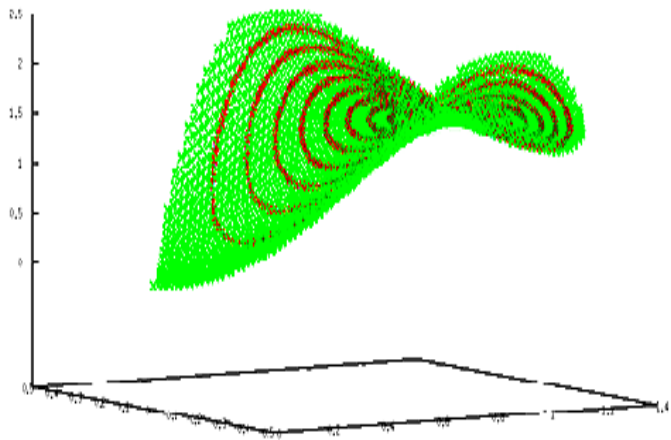}} 
\caption{  The extended probability density $\rho(x_r,x_i)$ for the $ A  <1$, $x_r>0$ region of the $n=1$ harmonic oscillator, evaluated along the trajectories, overlapped with the extended $\Psi^{\star}\Psi$ probability density. This probability density does not obey the continuity equation, but the total probability will be conserved. Also it agrees with the Born rule along the real line.}  \label{fig:n1shm_ls1}}    \end{figure}

Using the above definition $\rho^{\prime} \equiv \Psi^{\star}\Psi$ and the expression for $\dot{x}$ in the $n=1$ case, we shall see that

\begin{equation}
\frac{\partial (\rho^{\prime} \dot{x}_r)}{\partial x_r}+\frac{\partial (\rho^{\prime} \dot{x}_i)}{\partial x_i} \propto e^{-\alpha^2 (x^2-y^2)}(x_rx_i^3+x_r^3x_i). \label{eq:noncons_n1}
\end{equation}
The quantity on the right hand side is the density distribution of the `probability source'. It is positive for the first and third quadrants and negative for the second and fourth ones. When the  entire region inside the lemniscate is considered, the net source of probability is seen to be zero. Thus the total extended probability can be normalized for the $n=1$ harmonic oscillator.

It is found that the probability in the region inside the lemniscate is substantial;  $43.25\%$ of the total probability lies inside this region for the $n=1$ harmonic oscillator. The maximum value of $x_i$ for the lemniscate  (a measure of its width) in this case  is $x_i^{max}=  X_i^{max}/\alpha =0.4858/\alpha$. This explains how  classical particles are confined close to the real line. For instance, consider a classical harmonic oscillator of mass $m\sim 1$ kg and frequency $\sim 1$ Hz  in the $n=1$ state. The maximum value of $x_i$ for its lemniscate is $x_i^{max}\approx 10^{-17}/\sqrt{m\omega_0}$ in units of metres. Thus $43.25\%$ of the total probability in this case lies within the lemniscate of  width $\sim 10^{-17}$ m. It may also be noted that since the extended probability for $A>1$ decreases fast,  most of the probability outside  also lies close to the real line. 

\begin{figure}[ht] 
\centering{\resizebox {0.5 \textwidth} {0.3 \textheight }  
{\includegraphics {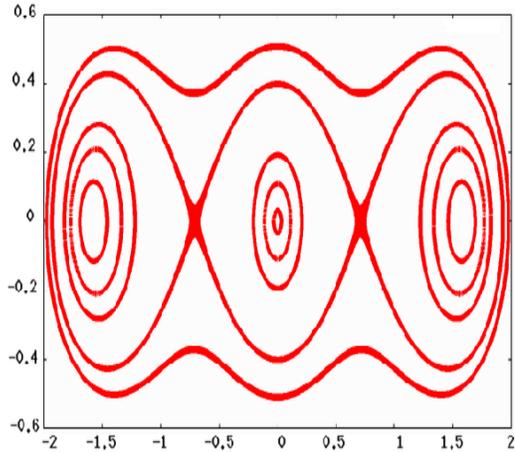}} 
\caption{ The complex trajectories in the $n=2$ harmonic oscillator case.}  \label{fig:n2shmtraj}} 
   \end{figure}

\begin{figure}[ht] 
\centering{\resizebox {0.5 \textwidth} {0.3 \textheight }  
{\includegraphics {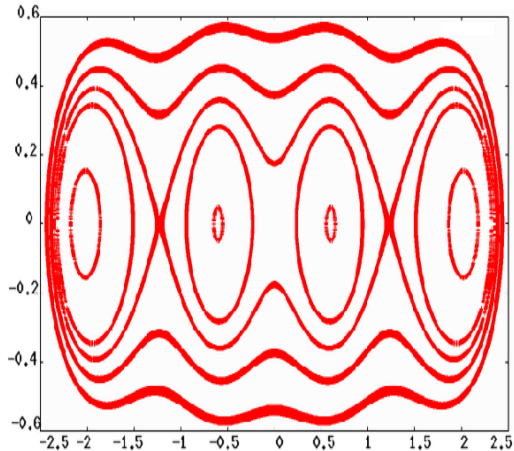}} 
\caption{ The complex trajectories in the $n=3$ harmonic oscillator case. }  \label{fig:n3shmtraj}} 
   \end{figure}

For the higher energy eigenstates of the harmonic oscillator,  the lemniscates are seen to be  narrower than that of the low energy ones. For example, in the $n=2$ case,  the complex paths are described by 

\begin{equation}
\left[ (X_r^2+X_i^2)^2-5(X_r^2-X_i^2)+\frac{25}{4}\right]^2(X_r^2+X_i^2) = \hbox{constant.}
\end{equation}
These are shown in Fig. (\ref{fig:n2shmtraj}). The  $X_i^{max}$ for the 3-fold lemniscate in this case is 0.4125 and hence its $x_i^{max}$  is less than that in the previous case. For  $n=3$, the complex paths are described by a more detailed expression, and are shown in Fig. (\ref{fig:n3shmtraj}). The $X_i^{max}$ for this 4-fold lemniscate  is found numerically as $\approx 0.39$, which is again  smaller than that in the previous cases. Thus it can be assured that for a classical oscillator of any energy and having mass 1 kg., the width of the region, where the large part of probability lies, is less than or of the order of $10^{-17}/\sqrt{\omega_0}$ m. This indicates that the probable imaginary part of position for classical particles are of  extremely small size. However, we may also see that  an electron executing harmonic oscillation with frequency $\omega_0$  has to accommodate relatively large values for $x_i$, approximately equal to $ 0.01/\sqrt{\omega_0}$ m. 

In summary, the probability axiom in the modified dBB quantum mechanics helps to distinguish  the classical limit of quantum harmonic oscillator as one in which the oscillator is probable to be found only  very close to the real axis. This  result is very important for complex quantum trajectories, for it explains  why the complex extension is not observable even indirectly in the classical limit.  We anticipate that this property is   true in other problems too.

\section{Conclusion}
The present work is a continuation of that in \cite{mvj2} to obtain a probability density for the entire complex $x$-plane, in the modified de Broglie-Bohm quantum mechanics. In the earlier work, a conserved probability (\ref{eq:rho_def}) was proposed  for particles   in 1-dimensional stationary states, but only for those regions  where  trajectories enclose poles of the velocity field. This continues to be so  in the present framework. But for those regions where trajectories do not enclose any poles, described as subnests, a conserved probability could not be found in \cite{mvj2} and it was considered  a missing link in the formalism. We here  adopt $\Psi^{\star}(x)\Psi(x)$, which is equivalent to that in (\ref{eq:rho_alt_ls1}), as the extended probability density for those special regions and note that the combined probability density  helps to find answers to many pertinent questions that arise in this context.

First of all, when compared to all other proposals for an extended probability density, the present combined probability has the important advantage that it is normalisable in the $x_rx_i$-plane. At the same time,  along the real axis in all regions, it agrees with the Born's probability, as required.  But we note that the proposed probability for subnests does not  obey a continuity equation. However, it can be seen that for the entire subnested region, the net value of this probability is conserved. This property of local nonconservation of particles, also found in the FFM trajectory representation, is argued to be characteristic of quantum phenomena in general and is not to be dismissed as unphysical. In addition, it explains why the imaginary part of complex trajectories do not leave any observable imprints in the case of classical particles. The present scheme is such that for a classical harmonic oscillator, probability is considerable only for those trajectories which lie extremely close to the real line, whereas for an electron in harmonic motion, trajectories have substantial probability even when their imaginary part  is appreciable.  The observation that  complex trajectories with large imaginary part are least probable in the classical limit is physically much relevant for the representation.

\end{document}